\documentclass[aps,prl,twocolumn,showpacs]{revtex4}
\usepackage{graphicx}
\usepackage{wasysym}

\begin{document}
\def\be{\begin{equation}}
\def\ee{\end{equation}}

\title{Generic features of the fluctuation dissipation relation
in coarsening systems}

\author{Federico Corberi}
\email{corberi@na.infn.it}
\affiliation{Istituto Nazionale per la Fisica della Materia,
Unit\`a di Salerno and Dipartimento di Fisica ``E. R. Caianiello'', 
Universit\`a di Salerno, 84081 Baronissi (Salerno), Italy}

\author{Claudio Castellano}
\email{castella@pil.phys.uniroma1.it}
\affiliation{Dipartimento di Fisica, Universit\`a di
Roma ``La Sapienza'', and Istituto Nazionale per la Fisica della Materia,
Unit\`a di Roma 1, P.le A. Moro 2, I-00185 Roma, Italy}

\author{Eugenio Lippiello}
\email{lippiello@sa.infn.it}
\affiliation{Istituto Nazionale per la Fisica della Materia,
Unit\`a di Salerno and Dipartimento di Fisica ``E. R. Caianiello'', 
Universit\`a di Salerno, 84081 Baronissi (Salerno), Italy}

\author{Marco Zannetti}
\email{zannetti@na.infn.it}
\affiliation{Istituto Nazionale per la Fisica della Materia,
Unit\`a di Salerno and Dipartimento di Fisica ``E. R. Caianiello'', 
Universit\`a di Salerno, 84081 Baronissi (Salerno), Italy}

\begin{abstract}

The integrated response function in phase-ordering systems with scalar,
vector, conserved and
non conserved order parameter is studied at various space dimensionalities.
Assuming scaling of the aging contribution
$\chi _{ag} (t,t_w)= t_w ^{-a_\chi} \widehat \chi (t/t_w)$
we obtain, by numerical simulations and analytical arguments,  
the phenomenological formula describing the dimensionality dependence
of $a_\chi$ in all cases considered.
The primary result is that $a_\chi$ vanishes continuously as $d$ approaches 
the lower critical dimensionality $d_L$. This implies that i) the existence of a
non trivial fluctuation
dissipation relation and ii) the failure of
the connection between statics and dynamics are generic features of
phase ordering at $d_L$.

\end{abstract}


\maketitle

After the groundbreaking work of Cugliandolo and Kurchan~\cite{Cugliandolo93}
on mean field spin glasses, 
the study of the out of equilibrium linear response function has been gaining
an  increasingly important role in the understanding of slow relaxation
phenomena. The key concept is that of the 
fluctuation dissipation relation (FDR)~\cite{Cugliandolo2002}. In terms of the
integrated response function $\chi(t,t_w)$, i.e. the response to an external
field acting in the
time interval $(t_w,t)$, an FDR arises  if 
$\chi(t,t_w)$ depends on time only through the unperturbed autocorrelation
function $C(t,t_w)$.
If this happens, there remains defined a function $\chi = S(C)$ which
generalizes the fluctuation dissipation theorem into the out of equilibrium
regime.

The existence of an FDR is important for several
reasons~\cite{Cugliandolo2002}.
Here we focus on a specific aspect: to what extent the FDR shape is
revealing of the mechanism of relaxation
and of the structure of the equilibrium state. In particular, we aim at
dispelling the common
belief that relaxation by coarsening and a simple equilibrium state do
{\it necessarily} imply a flat or trivial FDR, i.e. $S(C) = 1 - q_{EA}$
when $C$ falls below the Edwards-Anderson order parameter $q_{EA}$. 

To appreciate the relevance of
the problem, consider that, by reversing the argument, the observation of
a non flat FDR,would rule out  relaxation by coarsening.
This is a statement of far reaching consequences.
For instance, an argument of this type plays a role in the discrimination 
between the mean field and the droplet picture of the low temperature phase
of finite dimensional spin glasses~\cite{Ruiz}.
In that case, the final conclusion may well be right, but for the argument 
to be sound, the behavior of the response
function, when relaxation proceeds by coarsening, needs to be thoroughly
understood.

As a contribution in this direction, we have undertaken a large 
program of systematic investigation of the FDR in the context of phase
ordering systems~\cite{Bray94},
which provide the workbench for the study of all aspects of relaxation
driven by coarsening.
We have considered pure ferromagnetic systems quenched from above to below
the critical
point. We have covered the whole spectrum of systems with non conserved (NCOP),
conserved (COP), scalar ($N=1$) and vector ($N>1$) order parameter at different
space dimensionalities $d$, where $N$ is the number of components of the
order parameter. 
The manifold of the systems considered is displayed in Table I. Some of
these (marked by a dot) have been studied before.
The important novelty is that, with the new entries, the picture becomes rich enough 
to promote to
{\it generic} the behavior previously 
observed in the particular case of the Ising
model~\cite{Lippiello2000,Corberi2001,Castellano,Corberi2003}
and in the large $N$ model~\cite{Corberi2002}.
Namely, that a flat FDR is obtained for $d > d_L$, while a non flat FDR is
found for $d=d_L$, where
$d_L$ is the lower critical dimensionality. The implication is that a flat
FDR is not a necessary condition for coarsening.

To explain in more detail, let us recall~\cite{Cugliandolo2002} that,
quite generally, one can write
$\chi(t,t_w) = \chi_{st}(t -t_w) + \chi_{ag}(t,t_w)$.
The first is the stationary contribution due to the fast degrees of freedom
which rapidly equilibrate
with the bath, while the second is the aging contribution coming from
the slow out of
equilibrium degrees of freedom. What one can also show, in general, is that
a flat FDR is obtained if $\chi_{ag}(t,t_w)$ vanishes
asymptotically~\cite{Cugliandolo2002,Corberi2001}.
Now, in phase ordering for large $t_w$ one expects the scaling behavior
\be
\chi _{ag} (t,t_w)=
t_w ^{-a_\chi} \widehat{\chi} (t/t_w)
\label{1}
\ee 
from which it follows that the FDR is or is not flat according to $a_\chi >0$ or $a_\chi \leq 0$ 
(here we will restrict to the case with $a_\chi =0$).
Therefore, investigating the FDR shape requires the investigation of $a_\chi$. 

Let us see what is the situation with this exponent. 
In the Ising model~\cite{Lippiello2000,Corberi2001,Castellano,Corberi2003}
and in the large $N$ model~\cite{Corberi2002},  
we have found that  $a_\chi$
depends on dimensionality according to 
\be
a_{\chi}= \left \{ \begin{array}{ll}
  \delta \left ({d-d_L \over d_U-d_L} \right ) & \qquad $for$ \qquad d< d_U \\
  \delta  \qquad $with log corrections$ & \qquad $for$ \qquad d=d_U \\
  \delta & \qquad $for$ \qquad d > d_U 
\end{array}
\right .
\label{2}
\ee
where $\delta$ enters the time dependence of the density of defects
and $d_L$, $d_U$ are the two special dimensionalities
(with $d_L <d_U$) where $a_\chi=0$ and above which  $a_\chi =\delta$,
respectively. 
The density of defects goes like $\rho(t) \sim L(t)^{-n} \sim t^{-\delta}$,
where  $L(t) \sim t^{1/z}$ is the typical defect distance,
$z$ is the dynamic exponent  and $n=1$ or $n=2$ for scalar or vector
order parameter~\cite{Bray94}. Hence, $\delta= n/z$. 
 
Opposite to Eq.~(\ref{2}) stands a qualitative 
argument according to which
$\chi_{ag}(t,t_w)$ ought to be simply proportional to the defect
density~\cite{Barrat98,Franz98}. Namely, $a_{\chi} =\delta$ at any
dimensionality. There are no measurements or derivations of $a_{\chi}$
supporting such a statement. Furthermore, the argument is incompatible 
with the exact result $a_\chi =0$~\cite{Lippiello2000,Godreche2000} 
in the $d=1$ Ising model.
The implication for the FDR is that it
should be always flat, except for possible pathological cases.
This point of view is diametrally opposed to the one underlying
Eq.~(\ref{2}), where the vanishing of
$a_{\chi}$ at $d=d_L$, far from being a pathology, comes out
as a limiting behavior in the smooth dimensionality dependence
of $a_{\chi}$ for $d<d_U$.

The understanding of the FDR in coarsening systems requires to clarify
whether $a_{\chi}$ does or does not to depend on $d$.
Here we present strong evidence supporting Eq.~(\ref{2}) as
the generic pattern of behavior.

We have computed $\chi (t,t_w)$ for systems
quenched from infinite to zero final temperature. This
is computationally efficient and can be done without loss of generality,
since all
quenches below $T_C$ are controlled by the $T=0$ fixed point~\cite{Bray94}. 
In all cases we have used the time dependent Ginzburg-Landau
equation~\cite{Bray94}, except for NCOP with $N>1$ and $d>2$ where 
the Bray-Humayun~\cite{Bray90} algorithm has been used~\cite{nota2}.
The stationary response has been computed from equilibrium simulations. 
The aging part, then, has been obtained from
$\chi _{ag}(t,t_w)=\chi (t,t_w)-\chi _{st}(t-t_w)$.
To get $a_{\chi}$, one ought to extract the $t_w$
dependence of $\chi _{ag}(t,t_w)$ for fixed $x=t/t_w$~\cite{Corberi2003}.
However, this is computationally very demanding and would make it impossible
to get the vast overview we are aiming at. So, we have measured
$a_{\chi}$ from the large $t$ behavior for a fixed $t_w$, assuming
$\chi _{ag}(t,t_w) \sim t^{-a_{\chi}}$. This holds if 
$\widehat{\chi} (x) \sim x^{-a_{\chi}}$ for $x \gg 1$, which has been verified
in the NCOP scalar case~\cite{Corberi2001,Corberi2003} and it is an exact
result in the soluble models~\cite{Lippiello2000,Corberi2002}. The
assumption is that it holds in general.
The choice of $t_w$ is inessential provided it is larger than some
microscopic time necessary for scaling to set in~\cite{Corberi2003}.

\begin{figure}
\includegraphics[angle=0,width=7cm,clip]{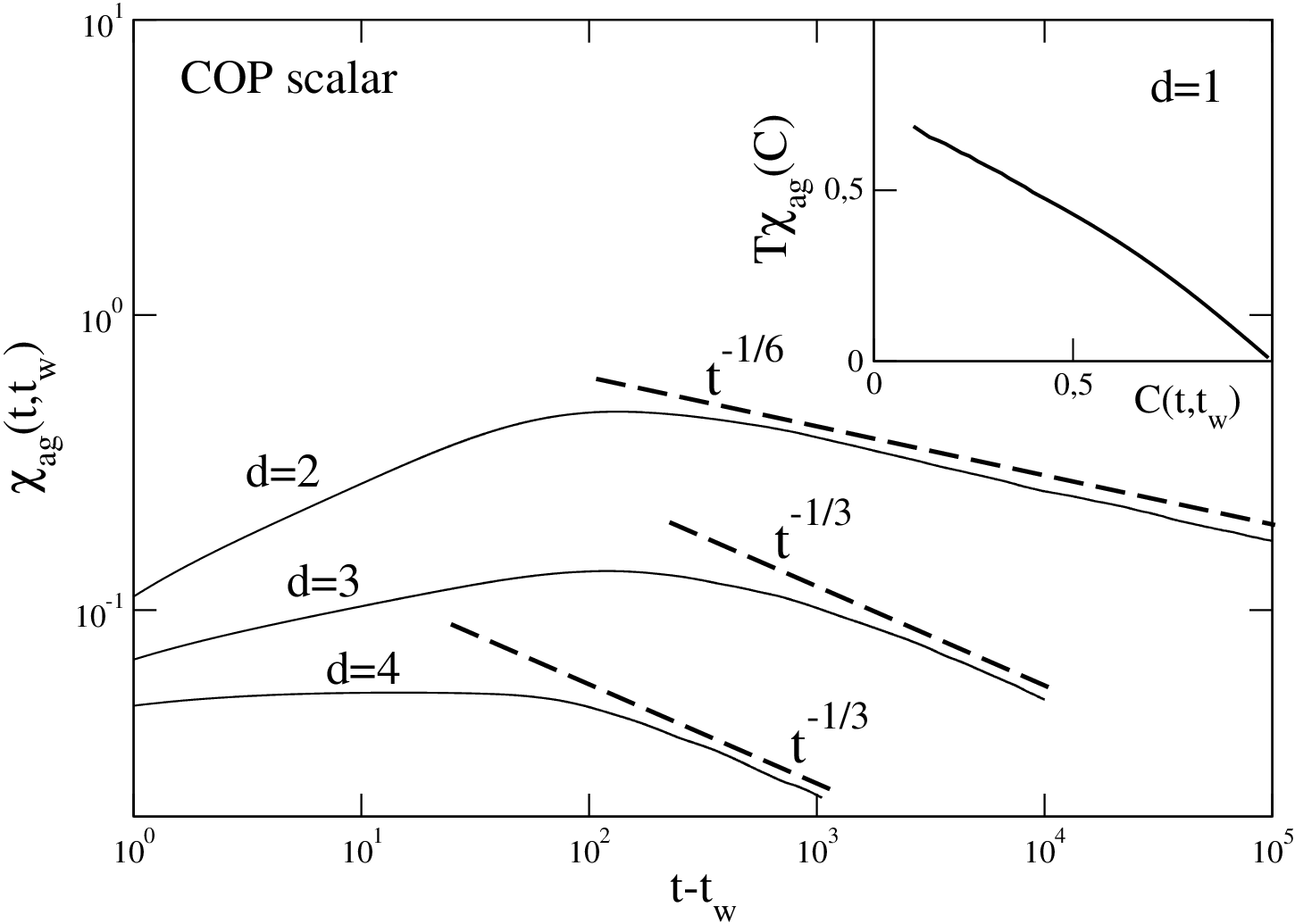}
\caption{$\chi _{ag} (t,t_w)$
against $t-t_w$ for $N=1$ with COP.
Lattice sizes, realizations and $t_w$: $512^2$, $41$ and $30$ for $d=2$;
$128^3$, $39$ and $40$ for $d=3$; 
$60^4$, $6$ and $31$ for $d=4$. The dashed lines are the slopes 
from Eq.~(\ref{2}). In the inset parametric plot for
$d=1$ from Ref.~\cite{Castellano}}.
\label{Fig1}
\end{figure}

\begin{figure}
\includegraphics[angle=0,width=7cm,clip]{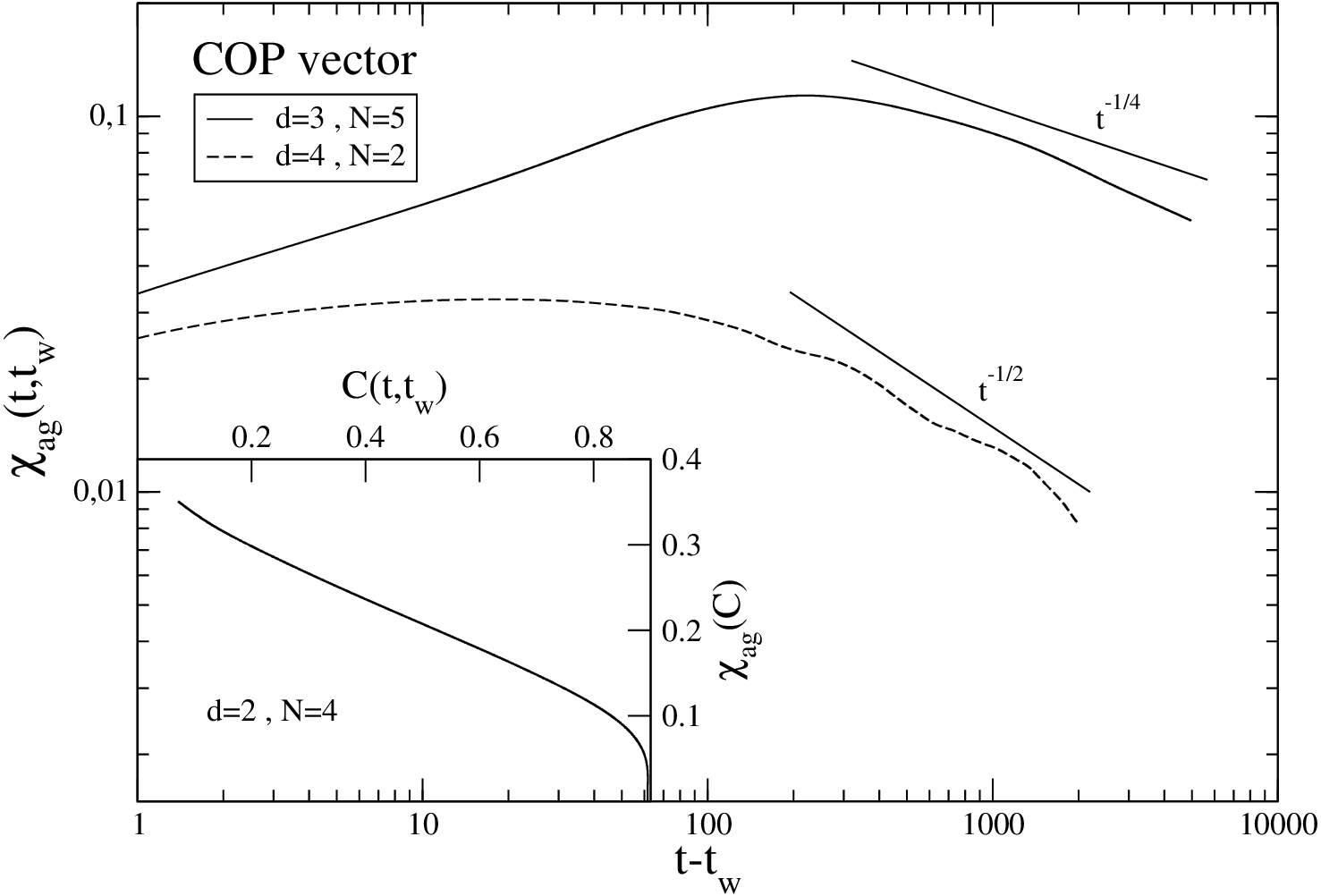}
\caption{$\chi _{ag} (t,t_w)$
against $t-t_w$ with COP.
Lattice sizes, realizations and $t_w$: $96^3$, $89$ and $35$ for $d=3$ and $N=5$;
$50^4$, $82$ and $35$ for $d=4$ and $N=2$. The dashed lines are the slopes 
from Eq.~(\ref{2}).
In the inset parametric plot for $d=2, N=4$.
Lattice size, realizations and $t_w$: $512^2$, $232$, $500$).}
\label{Fig2}
\end{figure}

\begin{figure}
\includegraphics[angle=0,width=7cm,clip]{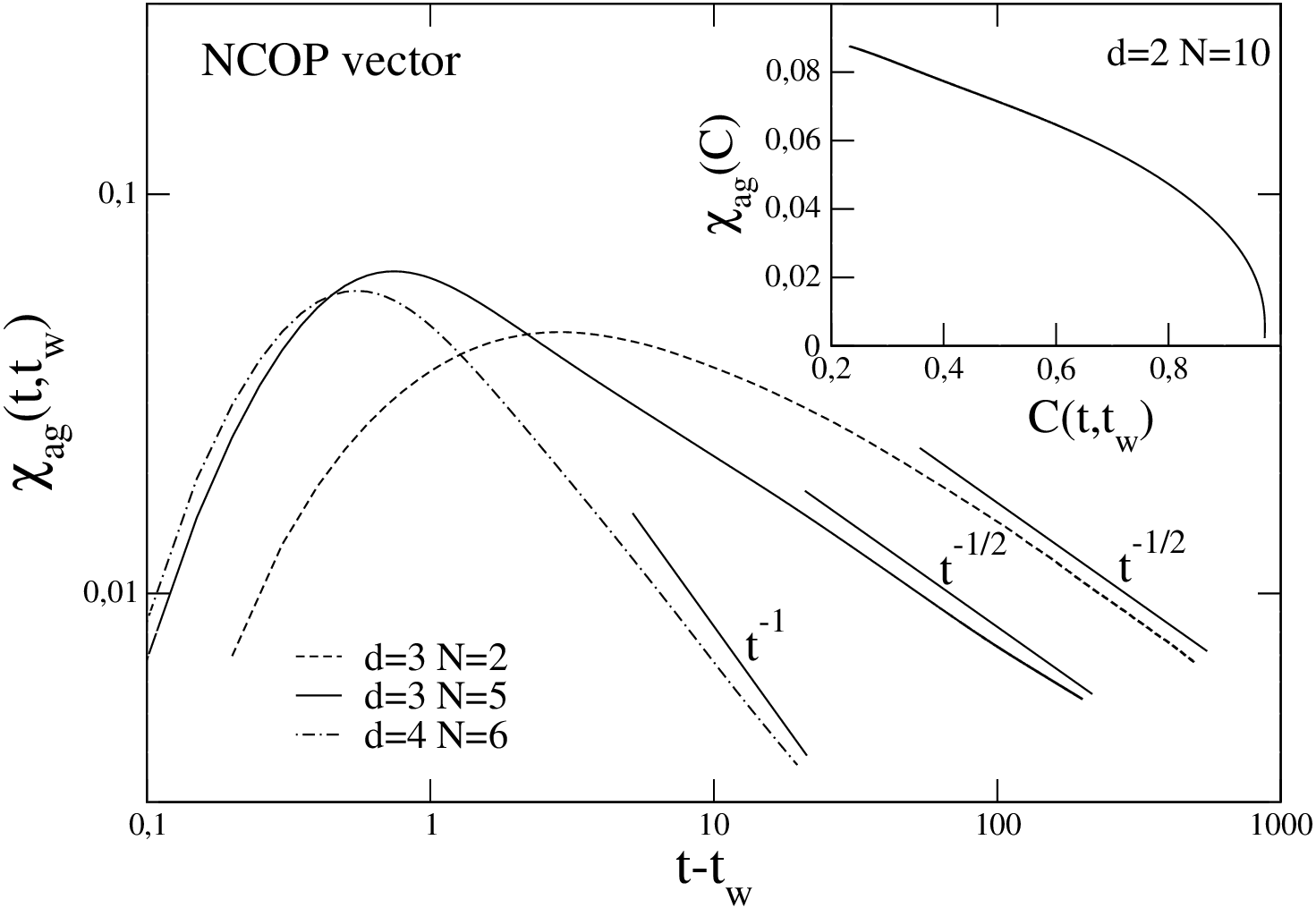}
\caption{$\chi _{ag} (t,t_w)$ 
against $t-t_w$ with NCOP.
Lattice sizes, realizations and $t_w$: $180^3$, $1445$ and $2$ for $d=3$ and $N=2$; 
$140^3$, $1486$ and $0.3$ for $d=3$ and $N=5$; 
$40^4$, $486$ and $0.3$ for $d=4$ and $N=6$. 
In the inset parametric plot for
$d=2$ and $N=10$. Lattice size, realizations and $t_w$: $1024^2$, $22$ and $20$).}
\label{Fig3}
\end{figure}

\begin{figure}
\includegraphics[angle=0,width=7cm,clip]{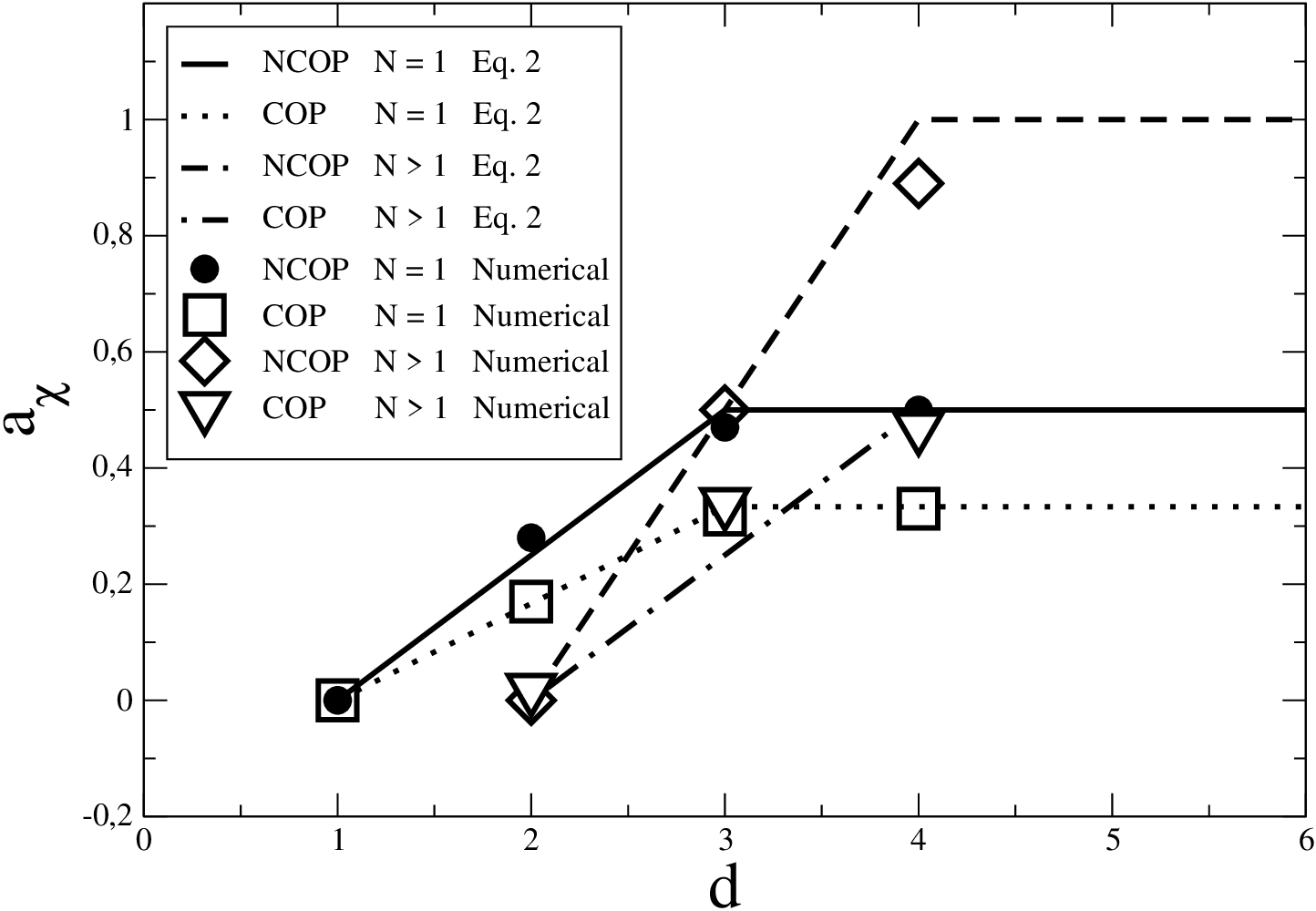}
\caption{Survey of results in table~\ref{Table2}.}
\label{Fig4}
\end{figure}

The time dependence of $\chi _{ag}(t,t_w)$ is depicted in
Figs.~\ref{Fig1},~\ref{Fig2} and~\ref{Fig3}.
We have extracted $a_{\chi}$ from the asymptotic power law decay and we
have collected all results, old and new,  in Table II. 
At $d_L$
we have used the parametric plot $\chi _{ag}(C)$ 
(insets of Figs.~\ref{Fig1},~\ref{Fig2} and~\ref{Fig3}), showing more effectively 
the absence of asymptotic decay, due to $a_{\chi}=0$. 
In Table II, we have also reported the
values of $a_{\chi}$ predicted by Eq.~(\ref{2}). 
The comparison with the
computed values is quite good. 
For convenience, we have collected in Table III the values of all the
parameters entering  Eq.~(\ref{2}).
Finally, Fig.~\ref{Fig4} provides the pictorial representation of Table II,
and it is the main result in the paper.

Let us now comment the results. From Fig.~\ref{Fig4} it is evident that the
pattern of behavior predicted
by  Eq.~(\ref{2}) is obeyed with good accuracy in the scalar cases,
with $d_L=1$ and $d_U=3$. 
In the vector cases, given the great numerical effort needed,  
values of $N$ were chosen according to the criterion of the best 
numerical efficiency,
together with the requirement to simulate both systems with ($N<d$) and without ($N>d$)
stable topological defects.
The overall behavior of the data in Fig.~\ref{Fig4} shows
that  Eq.~(\ref{2}) well represents the dimensionality dependence
of $a_\chi $ also in the vector case with $d_L=2$ and $d_U=4$.
Finally, the insets in Figs~\ref{Fig1},~\ref{Fig2} and~\ref{Fig3}
(together with the analogous figures for the $d=1$ Ising model in
Ref.s~\cite{Lippiello2000,Castellano} and in the large $N$ model~\cite{Corberi2002})
show quite clearly that $a_{\chi}=0$ and a non flat FDR are common features
in phase ordering kinetics at $d_L$.

At this stage Eq.~(\ref{2}) is a phenomenological formula.
Apart from the exact solution of the large $N$ model~\cite{Corberi2002},
there is no derivation of Eq.~(\ref{2}).
Here we propose an argument for the dependence of
$a_{\chi}$ on $d$ in the scalar case.
It is based on two simple physical ingredients:
{\bf a)} the aging response is given by the
density of defects $\rho(t)$ times the response of a single
defect~\cite{Corberi2001} $\chi_{ag}(t,t_w)=\rho(t)\chi_{ag}^s(t,t_w)$ and
{\bf b)} each defect responds to the perturbation
by optimizing its position with respect to the external field in a quasi-equilibrium way.
In $d=1$ this occurs via a displacement of the defect.
In higher dimensions, since defects are spatially extended,
the response is produced by a deformation of the defect shape.

\begin{figure}
\includegraphics[angle=0,width=7cm,clip]{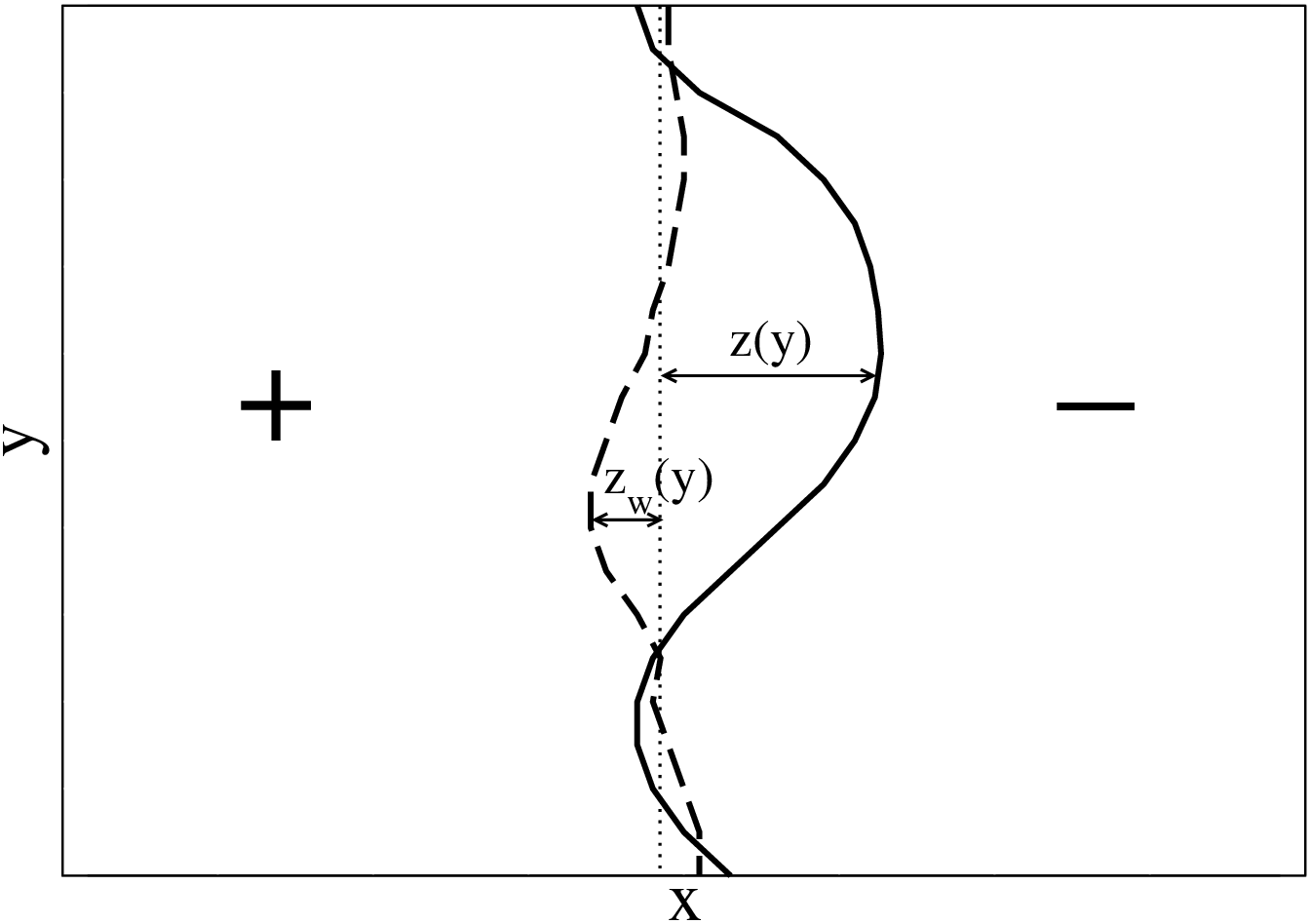}
\caption{Configurations with a single interface at time $t_w$ (dashed line)
and at time $t$ (continuous line).}
\label{Fig5}
\end{figure}

We develop the argument for a 2-d system,
the extension to arbitrary $d$ being straightforward.
A defect is a sharp interface separating two domains
of opposite magnetization. In order to analyse $\chi_{ag}^s(t,t_w)$ we consider
configurations with a single defect as depicted in Fig.~\ref{Fig5}.
The corresponding integrated response function 
reads~\cite{Corberi2001} 
$\chi_{ag}^s(t,t_w) =1/(h^2 {\cal L}^{d-1})\int dx dy \,
\overline{\langle S(x,y) \rangle h(x,y)}$, where $S(x,y)$ 
is the order parameter field which saturates to $\pm 1$ in the bulk of domains. 
$h(x,y)$ is the external random field with
expectations $\overline {h(x,y)}=0$,
$\overline {h(x,y) h(x',y')} =h^2 \delta(x-x') \delta(y-y')$ and 
${\cal L}$ is the linear system size.
The overbar and angular brackets denote averages over the
random field and thermal histories, respectively.
With an interface of shape $z_w(y)$ at time $t_w$ (Fig.~\ref{Fig5}), 
we can write 
$\chi_{ag}^s(t,t_w) = - 1/(h^2 {\cal L}^{d-1})
\overline{\int_{\{z \}} E_h \, P_h(\{z(y)\},t)}$,
where $P_h(\{z(y)\},t)$ is the probability that an interface profile $\{z(y)\}$
occurs at time $t$ and
$E_h = -\int _0 ^{\cal L} dy \int_{z_w(y)}^{z(y)} dx h(x,y) 
\mbox{sign} [z(y)-z_w(y)]$
is the magnetic energy. 
We now introduce assumption b) making the ansatz for the correction to the
unperturbed probability $P_0(\{z(y) \},t)$ in the form of a Boltzmann factor
$P_h(\{z(y)\},t)=P_0(\{z(y)\},t) \exp(-\beta E_h) \simeq P_0(\{z(y)\},t)
[1-\beta E_h]$. Then
$\chi_{ag}^s(t,t_w)=-1/(h^2 {\cal L}^{d-1})\overline {\int_{\{z \}} E_h(1-\beta E_h)
P_0(\{z(y)\},t)}$.
Taking into account that the term linear in $E_h$ vanishes by symmetry and 
neglecting $z_w(y)$ with respect to $z(y)$ for $t\gg t_w$, we eventually find 
$\beta ^{-1}\chi_{ag}^s(t,t_w)=
{\cal L}^{1-d} \int_{\{z \}} \int _0^{\cal L} dy
\vert z(y)\vert P_0(\{z(y)\},t)$. 
This defines a length which scales as
the roughness of the interface \cite{Henkel2003} given by
$W(t) = [{\cal L}^{1-d} \int_{\{z \}} \int dy z(y)^2 
P_0(\{z(y)\},t)]^{1/2}$.
The behavior of $W(t)$ in the coarsening process can be inferred
from an argument due to Villain~\cite{Abraham89}.
In the case $d\le 3$, when interfaces are
rough~\cite{Rough}, for NCOP one has $W(t) \sim t^{(3-d)/4}$, while for
COP $W(t) \sim t^{(3-d)/6}$, with logarithmic corrections in both cases 
for $d=3$. 
For $d>3$ interfaces are flat and $W(t)\simeq const.$
Finally, multiplying $\chi_{ag}^s$ by 
$\rho (t) \sim L(t)^{-1}$ 
Eq.~(\ref{2}) is recovered~\cite{note2} and 
$d_U$ is identified with the roughening dimensionality $d_R=3$.
The relevance of roughening in the large time behavior of $\chi (t,t_w)$
has been independently pointed out by Henkel, Paessens and Pleimling
in Ref.~\cite{Henkel2003}. The crucial difference with these authors is that they 
believe roughening to be unrelated to aging behavior, while we claim the opposite.

In summary, we have investigated the scaling properties of the response
function over 
a large variety of systems designed to bring forward the generic features when
relaxation is driven by coarsening. The primary result is that the exponent 
$a_{\chi}$ depends on dimensionality and that it vanishes smoothly as
$d \rightarrow d_L$.
This implies that a non trivial FDR is not exceptional, rather is the rule
for  coarsening systems at $d_L$. Another important consequence is that the
failure of the connection between statics and dynamics at
$d_L$~\cite{Corberi2001} is also a generic feature of coarsening.
The connection
between the FDR and the overlap probability function is derived~\cite{Franz98}
under the assumptions of stochastic stability and that $\chi(t,t_w)$ goes
to the equilibrium value
as $t \rightarrow \infty$. The latter assumption does not hold at $d_L$
due to the existence of a non flat FDR 
(insets of Figs~\ref{Fig1},\ref{Fig2},\ref{Fig3}), which makes the limiting 
value of $\chi(t,t_w)$ to rise above the equilibrium value. Obviously,
the important and, as of yet, unanswered question is why all this happens
at $d_L$. The scaling behavior of the response function reported in
this Letter adds to the many already existing challenges posed by a theory
of phase ordering kinetics.

This work has been partially supported from MURST through PRIN-2002.

\begin{table}
\begin{ruledtabular}
\begin{tabular}{||c|c|c||c|c||}
    & \multicolumn{2}{c||}{NCOP}& \multicolumn{2}{c||}{COP}\\ \hline \hline
$d$ & $N=1$      & $N>1$        & $N=1$     & $N>1$     \\ \hline
1   & $\bullet$  &              & $\bullet$ &           \\ \hline
2   & $\bullet$  & $N=10$        & $N=1$     & $N=4$      \\ \hline
3   & $\bullet$  & $N=2$, $N=5$ & $N=1$     & $N=5$     \\ \hline
4   & $\bullet$  & $N=6$        & $N=1$     &           \\ 

\end{tabular}
\end{ruledtabular}
\caption{The manifold of systems considered. Entries with dots correspond 
to systems 
studied in Refs.~\cite{Lippiello2000,Corberi2001,Castellano,Corberi2003}.}
\label{Table1}
\end{table}

\begin{table}
\begin{ruledtabular}
\begin{tabular}{||c|c|c|c|c||c|c|c|c||}
    & \multicolumn{4}{c||}{$N=1$}& \multicolumn{4}{c||}{$N>1$}\\ \hline
    & \multicolumn{2}{c|}{NCOP} 
    & \multicolumn{2}{c||}{COP} 
    & \multicolumn{2}{c|}{NCOP} 
    & \multicolumn{2}{c||}{COP} \\ \hline \hline
$d$ & Eq.~(\ref{2}) & Best fit & Eq.~(\ref{2}). & Best fit & Eq.~(\ref{2}) & Best fit & Eq.~(\ref{2}) & Best fit
\\ \hline
1   & 0 & 0 $\bullet$ & 0 & 0 $\bullet$& & & & \\ \hline
2   & 1/4 & 0.28 $\bullet$ & 1/6 & 0.17 & 0 & -0.07 & 0 & -0.13 \\ \hline
3   & 1/2 $(\log)$ & 0.47 $\bullet$ & 1/3 $(\log)$ & 0.32 & 1/2 & 0.50 & 1/4 & 0.34 \\ \hline
4   & 1/2 & 0.50 $\bullet $& 1/3 & 0.33 & 1 $(\log)$ & 0.89 & 1/2 $(\log)$ & 0.47 \\ 

\end{tabular}
\end{ruledtabular}
\caption{Exponent $a_\chi$ from Eq.~(\ref{2}) and from best fit of numerical data.
Values marked by a dot come from 
Refs.~\cite{Lippiello2000,Corberi2001,Castellano,Corberi2003}.}
\label{Table2}
\end{table}

\begin{table}
\begin{ruledtabular}
\begin{tabular}{||c|c|c|c|c||}
    & \multicolumn{2}{c|}{$N=1$}& \multicolumn{2}{c|}{$N>1$}\\ \hline
           & NCOP & COP & NCOP & COP \\ \hline
z          & 2    & 3   & 2    & 4   \\ \hline
$\delta$   & 1/2  & 1/3 & 1    & 1/2 \\ \hline
$d_L$  &  \multicolumn{2}{c|}{1} & \multicolumn{2}{c|}{2} \\ \hline
$d_U$  &  \multicolumn{2}{c|}{3} & \multicolumn{2}{c|}{4} \\

\end{tabular}
\end{ruledtabular}
\caption{Parameters entering Eq.~(\ref{2}).}
\label{Table3}
\end{table}

\begin{thebibliography}{99}
\bibitem{nota2}
The algorithm by Bray and Humayun enters the scaling regime earlier
than the GL equations, so that a sufficiently long scaling window exists
before finite size effects appear.
\end{thebibliography}
\end{document}